\documentclass{sig-alternate}

\usepackage{makeidx}  
\usepackage{times,amsmath,epsfig}
\usepackage{epstopdf}
\usepackage{amsmath}
\usepackage{listings}  
\usepackage{color}
\usepackage{rotating}
\usepackage{enumerate}
\usepackage{url}
\usepackage{enumitem}
\usepackage{multirow}
\usepackage{tabu}
\usepackage{amssymb}
\usepackage{graphicx}
\usepackage{float}
\usepackage{hyperref}
\usepackage[utf8]{inputenc}

\newtheorem{defn}{Definition}

\urldef{\mailNUIG}\path|firstname.lastname@iai.uni-bonn.de|

\lstdefinelanguage{turtle} 
{morekeywords={@prefix}, 
sensitive=false, 
morecomment=[l]{\#}, 
}
\newcounter{usecaseno}
\DeclareRobustCommand{\usecase}[1]{
   UP\refstepcounter{usecaseno}\theusecaseno\label{#1}}

\newcounter{qano}
\DeclareRobustCommand{\ca}[1]{
   CA\refstepcounter{qano}\theqano\label{#1}}

\begin{document}
%

\title{"How much?" Is Not Enough\\ An Analysis of Open Budget Initiatives}

%
%
%
%
%

\numberofauthors{5} 
%
\author{
\alignauthor
Alan Tygel\\
       \affaddr{Graduate Program on Informatics -- PPGI -- UFRJ, Brazil}\\
       \email{alantygel@ppgi.ufrj.br}
\alignauthor
Judie Attard\\
       \affaddr{University of Bonn, Germany}\\
       \email{attard@iai.uni-bonn.de} 
\alignauthor
Fabrizio Orlandi\\
       \affaddr{University of Bonn, Germany}\\
       \email{orlandi@iai.uni-bonn.de}      
\and
\alignauthor
Maria Luiza Machado Campos\\
       \affaddr{Graduate Program on Informatics -- PPGI -- UFRJ, Brazil}\\
       \email{mluiza@ppgi.ufrj.br}      
\alignauthor
S\"{o}ren Auer\\
       \affaddr{University of Bonn and Fraunhofer IAIS, Germany}\\
       \email{auer@cs.uni-bonn.de}
}








\maketitle
\begin{abstract}
A worldwide movement towards the publication of Open Government Data is taking place, and budget data is one of the key elements pushing this trend. Its importance is mostly related to transparency, but publishing budget data, combined with other actions, can also improve democratic participation, allow comparative analysis of governments and boost data-driven business. However, the lack of standards and common evaluation criteria still hinders the development of appropriate tools and the materialization of the appointed benefits. In this paper, we present a model to analyse government initiatives to publish budget data. We identify the main features of these initiatives with a double objective: (i) to drive a structured analysis, relating some dimensions to their possible impacts, and (ii) to derive characterization attributes to compare initiatives based on each dimension. We define use perspectives and analyse some initiatives using this model. We conclude that, in order to favour use perspectives, special attention must be given to user feedback, semantics standards and linking possibilities.
\end{abstract}

\terms{open government data, open budget initiatives, e-government, participation, transparency}

\section{Introduction}
\label{sec:introduction} 
In the last six years, a worldwide movement towards the publication of Open Government Data (OGD) has been taking place. 
The aims and scope of OGD initiatives in each country are diverse and we can count almost 100 countries publishing some kind of OGD\footnote{According to the Open Data Index: \url{http://index.okfn.org/}.}. 

The motivation for governments to publish OGD are also diverse.
It ranges from the democratic point of view with increasing government transparency and citizen participation to the more economic motivation of fostering new data-driven businesses. 
The strengthening of law enforcement has also fostered OGD publishing~\cite{Huijboom2011}.

A large number of stakeholders may take part of the OGD ecosystem, namely: as data providers, different levels of public administrations (including local, regional, national and transnational), and citizens, and as consumers, civil society initiatives and NGOs, companies, journalists and media organisations. 
While data providers mostly play the specific role of publishing data in an open format, other stakeholders participate in this initiative in a number of ways, including viewing the open data, sharing feedback, and exporting data into their own systems.
It is also expected that these stakeholders behave as \emph{prosumers}, not only passively consuming data, but also interfering in its production and publication.

OGD can be related to a diversity of themes. 
Education, crime, health, transportation and company registration are common subjects. 
However, one type of data is of particular importance: government budgetary data, as timely access to these data is critical to accomplish government accountability.   

All governments and public administrations maintain budgetary data, unlike, for example, bus position data, which depends on sensors, or data about the occurrence of a specific disease, which depends on a health information system.
From the citizen side, information on budget is a key element to ensure that public funds are being properly used.
In locations where a participatory budget was implemented, that is, part of the budget allocation is decided by the community, access to this kind of data is indispensable.
A global initiative to improve openness of governments -- the Open Government Partnership (OGP) -- has the fiscal transparency as a minimum eligibility criteria\footnote{Other criteria can be found at \url{http://bit.ly/1929F1l}.}, characterizing budget data as a foundation of open government.


Even with so many possible positive impacts, existing public financial transparency portals suffer from a number of shortcomings.
First of all, they suffer from the large number of diverse data structures that make the comparison and aggregate analysis of transnational financial flows practically impossible. 
The tools to present, search, download and visualise this financial data are also nearly as diverse as the number of existing portals. 
This heterogeneity~\cite{Vafopoulos2013} may even prevent an analysis of the quality of the data for the same funds administered by different funding authorities. 
Past efforts have sought to overcome this situation by creating comprehensive and connected transparency portals, such as Farmsubsidy.org, and more recently, Publicspending.net.

Within the existing open budget initiatives, low user engagement has been reported~\cite{Worthy2013}. 
Moreover, most of the budget publishing efforts result in simple data catalogues, fragmented and dispersed, because they do not share standards and methodologies~\cite{Vafopoulos2013}. 
The absence of standards can lead to data misuse~\cite{Zuiderwijk2014a}, or even to results opposed to the initial aims~\cite{Gurstein2011}.

The basis for such standards has to be set.
Together with other ongoing initiatives~\cite{OpenSpending, Vlasov2014}, we believe that the development of a solid standard can help governments to make their budget data more usable, and thus enable citizen participation in the democratic process.
In this article we define a structured analysis framework for budget data, which can help developers and policy makers to understand the importance of various aspects of budget data publishing and to develop more adequate budget publishing systems. 
After defining some foundational concepts (Section~\ref{sec:openbudgetdata}), highlighting the importance of budget data (Section~\ref{sec:whybudgetdata}) and discussing related work (Section~\ref{sec:relworks}), we describe the chosen methodology (Section~\ref{sec:methodology}) and derive dimensions and characterization attributes, based on three use perspectives (Section~\ref{sec:framework}).
These characterization attributes are applied to 23 open budget initiatives (Section~\ref{sec:analysis}), and results are discussed (Section~\ref{sec:conclusions}).

\section{What is Budget Data?}
\label{sec:openbudgetdata} 

Open Budget Data is the topic of a few recent publications~\cite{Vlasov2014, Vafopoulos2014, Martin2013, OpenSpending2014, Chambers2012}. 
Nevertheless, it is important to establish a common ground to some basic concepts, as they have not always a single widely accepted definition.
Here, we propose definitions for \emph{Budget}, \emph{Spending} and \emph{Revenue}, as the main quantities tackled, and \emph{Open Budget Data} as the general term\footnote{A further discussion about these terms can be found in \url{http://community.openspending.org/research/handbook/types-of-spending-data/}.}.

\begin{defn}
\textbf{Budget} is the description of the amount of money planned to be spent in a specified time period. 
Budget descriptions can refer to several levels of specificity, from general (total amount to be spent) to specific (amount by area, or category).
A budget description can be characterized by: 
\begin{itemize}
	\item (i) the scope, that is, the corresponding administrative level (municipality, region, country etc.); 
	\item (ii) Optionally, a domain, such as healthcare, public transportation; 
	\item (iii) if applicable, the related location (region, city, neighbourhood, or latitude and longitude) and 
	\item (iv) a period of time.
\end{itemize}
Budget comprises a set of budget items which have a budget category and an associated amount with a currency. 
Categories can be organized hierarchically, where higher levels of the hierarchy are representing aggregations of the lower levels.
There are different types of budget, such as proposed, planned, and certified, which is presented after the budget term. 
Budgets may also receive amendments during their associated term.
\end{defn}

\begin{defn}
\textbf{Spending}, or expenditure, refers to the amount of money actually spent by the public administration.
It can also be seen as the realisation of the budget.
Government spending can be split in four main categories: 
\begin{itemize}
	\item (i) Transfer payments, related to social benefits as pension, housing, or floor income for low income households; 
	\item (ii) Current government spending, related to the costs of maintaining the government structure, mainly public employees salaries;
	\item (iii) Capital spending, which goes for building infrastucture, as roads, hospitals, schools etc; and
	\item (iv) Financial costs, as internal and external debt services.
\end{itemize}
Ideally, spending should be published in the finest grain: transactions, which is the description of every payment, including value, time period and recipient. 
Transactions should also be classified according to properly defined criteria to generate aggregate amounts. 
These criteria are the same as specified for the budget: scope, domain, place and time.
There exists also different types of spending, such as planned (according to the budget), authorized (payment order) and executed (money transferred from government to the recipient).
\end{defn}

\begin{defn}
\textbf{Revenue} is the amount of money received by a government administration. 
Revenues can have several types of origins, such as taxes (revenue, commercialization), service fees (transportation), royalties (oil and mine exploration), concessions (roads, electromagnetic spectrum) or financial operations. 
Predicted revenues, used to specify the budget may differ from the actual revenues.
\end{defn}

\begin{defn}
\textbf{Open Budget Initiative} refers to any portal or application which publishes budget, spending and/or revenue data, that allows the civil society -- IT experts or not -- to access those data. 
It may comprise one or many datasets, which can be downloaded in several formats or directly visualized in tables, charts or maps.
The model presented in Section~\ref{sec:framework} describe an Open Budget Initiative in further details.
\end{defn}

\section{Why Budget Data?}
\label{sec:whybudgetdata}

The importance of publishing government budgetary data can be summarised in five key elements:

\hspace{.8cm}

\noindent\textbf{Transparency}: Opening budget data unveils public funds management. 
This increases accountability and therefore augments citizen's trust in public administration, whilst having a potential of uncovering hidden transactions and thus preventing corruption. 
An important factor which can stimulate corruption is the fact that funding goes through the hands of public officials without further scrutiny. 
In European Union Member States, this is particularly evident within public procurement, which is prone to corruption owed to deficient control
mechanisms~\cite{EuropeanComission2014}.
Essentially, such acts are concealed from the public eye. 
Supporting financial transparency enhances accountability within public sectors and, as a result, prevents corruption. 

\hspace{.8cm}

\noindent\textbf{Participation}: Opaque regimes may compel citizens to engage against the government. 
A transparent public administration, on the contrary, can stimulate social participation in community enhancement.
Open budget initiatives can not only enable meaningful civil society scrutiny of transnational financial flows, but they can also provide platforms for stakeholders to develop benchmarks that create pressure on public authorities to provide data in a timely, comparable, re-useable and well-structured fashion. 
These platforms can also involve local citizens in the budget planning and auditing phases, by allowing them to interact with the process, providing opinions and suggestions on setting budget priorities, providing feedback on the published transactions.
A virtuous circle can be created, in which both public officials and civil society will realise the value of data and analysis tools, in a collaborative environment open to contributions and engagement. 

\hspace{.8cm}

\noindent\textbf{Comparative Analysis}: Well organized budget data facilitates researchers and policy makers to compare spending strategies between cities, states and countries, and also among different administration levels.
Visualisation, analytics and exploration tools can offer different stakeholders an opportunity to scrutinize and interpret financial data related to a region of interest. 
It also allows to compare allocations and transactions between multiple regions, to visualise detected trends and budget projections and to investigate anomalies and activities, which have been flagged as suspicious.
A necessary condition for that is the compatibility and consistency of data from different data sources. 

\hspace{.8cm}

\noindent\textbf{Efficiency and Effectiveness}: Efficiency of public spending can be assessed by comparing, for example, the cost per kilometre of a railway. 
The effectiveness can also be assessed, in this case, by the revenues generated with the railway.

\hspace{.8cm}

\noindent\textbf{Business Value}: 
It has been recently stated that "Open data can help unlock U\$3 trillion to U\$5 trillion in economic value annually"~\cite{Manyika2013}.
Publishing budget data can stimulate the creation, delivery and use of new services on a variety of devices, utilising new web technologies, coupled with open public data. 
These services include visualisation services and data discovery services, such as data mining and comparative analysis, which enable stakeholders to explore the data, identify patterns, as well as potentially forecasting budget and transaction trends. 
Budget data can also generate value by empowering journalists when they report on spending items.
Accurate information on public funds usage may enable content producers to create better articles.

\section{Related Works}
\label{sec:relworks}

A number of recent works proposed frameworks, impact measures or comparison criteria on the general open data domain. 
Some of them aim the comparison of e-government and open data policies \cite{Zuiderwijk2014, Veljkovic2014}. 
In~\cite{Ubaldi2013}, a framework is proposed to evaluate OGD initiatives, pointing also to the development of impact metrics. 

A theoretical background to analyse the impact of OGD was developed in~\cite{Granickas2013}. 
Impacts are divided into economical, political and social, and for each of them, possible implementation issues and impact metrics are deeply discussed. 
Recently, a working group was created to develop methods for assessing open data. In their first report~\cite{Caplan2014}, a draft of a framework is proposed.

Automatic benchmarking techniques are proposed in~\cite{Atz2015}. 
Despite enabling large scale with low cost and high frequency evaluations, automatic assessment can miss some political and social aspects of open data.

Even though structured analysis and comparison of open budget initiatives have not received much attention from the literature, two works must be highlighted. 
The Open Budget Survey~\cite{InternationalBudgetPartnership2012} is a research project that, every two years, "measures the state of budget transparency, participation, and oversight in countries around the world". 
It generates the Open Budget Index\footnote{\url{http://www.obstracker.org/}}, which is updated monthly, and is based on the publication of eight key budget documents. Despite being a very useful comparison tool, this methodology does not evaluate information systems used to publish budget data, which are the way how the information reaches the society.

An evaluation and comparison between almost 30 Brazilian government transparency portals, on several administration levels, is presented in~\cite{Beghin2014}. 
The analysis was based on the 8 Open Government Principles\footnote{\url{http://opengovdata.org/}} evaluated for each portal by experts. 
Despite being a well defined and wide accepted model, these principles are quite general, and do not refer to specific characteristics of budget data. 
Moreover, they cover basically the publisher side.

\section{Methodology}
\label{sec:methodology} 
The research approach used to develop this model was inspired by the observation, induction and deduction method used in~\cite{Zuiderwijk2014}. 
After analysing the related bibliography and observing some randomly collected open budget initiatives, we used an inductive reasoning to build the first approach to the model.
The model is a set of \emph{dimensions}, which represent different themes to be assessed in an open budget initiative. Dimensions are grouped in \emph{parts}, according to its general functions.

The same basis is used to define \emph{use perspectives} (UP), which represent different ways of using budget data. 
From the UPs, we extract related requirements.

Model and use perspectives were then applied to other open budget initiatives in a deductive reasoning, in order to verify the fitness of the dimensions, and the coverage of the use perspectives.
Missing items were added to the model and to the use perspectives, and the feedback loop was run until no significant changes were found.
Finally, use perspectives were checked against the model, in order to verify the correspondence between model dimensions and use perspectives.
This correspondence is materialized in the \emph{characterization attributes}

The result of this observation, induction and deduction approach is described in the next section.

\section{A model to analyse open government budget data portals}
\label{sec:framework} 

\begin{figure*}[ht]
\begin{center}
\includegraphics[scale=0.75]{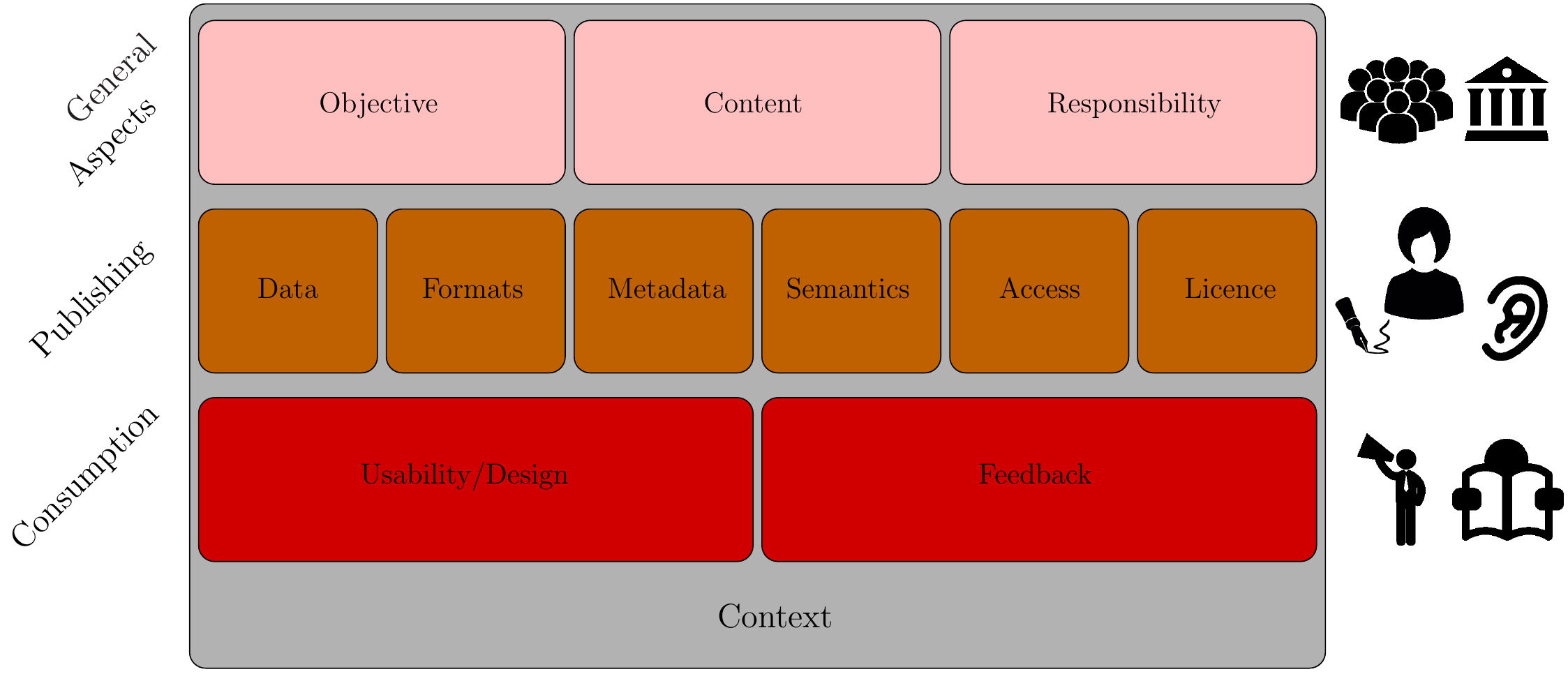}
\caption{Model to analyse open budget initiatives. The four parts -- General Aspects, Publishing, Consumption and Context -- are interconnected, and composed by several dimensions. Icons made by \href{http://www.flaticon.com/authors/icomoon}{Flaticon (CC)}.}
\label{fig:framework}
\end{center}
\end{figure*}

The main objective for building this model is the need for a mechanism to assess different strategies for publishing budget data. 
We do not aim to build rankings, but rather to systematize open budget initiatives in order to assess their fitness to specific use perspectives. 
A general overview of the proposed model is depicted in Figure~\ref{fig:framework}. 
The model consists of four parts: 
\begin{enumerate}
\item \emph{Context}, referring to external aspects related with the initiative;
\item \emph{General Aspects}, referring to the overall characterization of the initiative;
\item \emph{Data Publishing}, referring to aspects specific to data publishing process; and
\item \emph{Data Consumption}, referring to aspects specific to the data consumption process.
\end{enumerate}

Naturally, there is a strong coupling between these parts. 
The way data are published affects directly the consumption. 
By the same reasoning, the feedback generated by users (should) affect data publishing. 
The context particularly impacts the general aspects, but also influences the other parts.

The context part represents the environment in which the open budget initiative is involved. 
It stands for the open data policies and legislation which rules the publication of spending data, and also the government initiatives to promote the use of data, either only by advertising, or more incisively promoting data literacy.
Although we recognize that the context is a key element for the success of an open budget initiative, we will not consider it in the scope of this paper because its complexity would make the first approach to an objective model unfeasible. 
For the time being, we will focus on the general aspects directly related to the initiative, and on the issues related to publishing and consuming data. 

Each part of the model is composed of several dimensions, which will be assessed through \emph{Characterization Attributes}:

\begin{defn}
\emph{Characterization Attributes} are features of open budget initiatives that: 
(i) are objectively assessable; 
(ii) expect qualitative values; and 
(iii) have direct impact on the realisation of use perspectives.
\end{defn}
The characterization attributes derived from the dimensions are summarised in Table~\ref{tab:characterizationattributes}.

Characterizing an open budget initiative is the first step in order be able to assess quality. 
The term quality may refer to different concepts. 
In this work, we define quality as the conformance to requirements, which in our case are those associated to the use perspectives. In other words, we can say that quality is the fitness for use. Thus, we define three use perspectives, from which we extract some requirements:

\hspace{.8cm}

\noindent\textbf{\usecase{up:transp} -- Transparency:} Journalists, software developers, NGOs, and grass-roots movements use budget data to audit government and to translate data into more accessible formats for the society. For this use case, detailed data (i.e., transaction level), consistent classification levels, and machine readable formats are some important requirements.
Discussion and feedback on the provided data are also requirements in this case, for example, for suggesting different priorities for budgeting, or discussing a particular transaction. 
Both citizens and public administration benefit from this feature since the citizens (or other stakeholders) can show their perspectives and the public administration entity would check the current priorities to see if they need to be amended. 

\hspace{.8cm}

\noindent\textbf{\usecase{up:part} -- Participation:} 
For the last two decades, cities from all over the world have been implementing participatory budgeting (PB) experiences with different systems and procedures. Research shows how developing and promoting PB digital solutions can increase civic engagement up to seven times~\cite{Sintomer2008}. 
In Europe, digital solutions to promote citizen engagement in budget creation include, for example, sending proposals by email, participating in online forums and discussion, subscription to SMS updates and video streaming~\cite{Peixoto2009}. 
Germany presents one of the most advanced digital solutions to engage citizens, as shown in the participatory budgeting portal of the city of Freiburg\footnote{\url{http://www.beteiligungshaushalt-freiburg.de}}.
The participation use perspective will be exemplified by the PB case.
PB members must have access to accurate and easily understandable budget data. 
Through this perspective, design, usability, and human readable formats are the most important requirements.
Hierarchically aggregated categories also play an important role.

\hspace{.8cm}

\noindent\textbf{\usecase{up:policy} -- Policy Making:} If adequately published, budget data can be used to compare the way each government manages public funds. 
Researchers and policy makers should be able to compare the budgets and spending data between (i) different public administrations (e.g. Cologne vs Münich); or (ii) different periods (e.g. year 2013 vs year 2014),
 and thus relate spending strategies to political, economical and social outcomes. 
Comparing spending profiles among governments requires the use of common classifications, vocabularies and ontologies, and the possibility of linking data with other databases, as, for example, multinational enterprises data~\cite{Vafopoulos2013}.
In order to enable the integration of the corresponding budget data on the different public administration contexts, a semantic data model for budgets and spending has to be defined.
In this case, publishing financial data in a reusable, machine-processable, linked-data format can enable integration and reuse across multiple sources. 
The use of a standard format also facilitates the comparison of data from different municipalities or regions. 
More importantly, it allows all the stakeholders involved or interested in budget planning or spending, to manipulate data using the same tools and methods, thus supporting financial transparency in public budgeting and spending.
This may allow the creation of visualisations and comparative data analyses for the discovery of trends. 
Stakeholders will therefore be able to view and compare allocated budgets and transactions, and give feedback on each item. 
This feedback can then be shared through social media and also be directly exploited by governments and public administrations to achieve better budget management. 
The latter two stakeholders will thus benefit from receiving targeted suggestions, comparative benchmarks and scenarios.

\hspace{.8cm}

In the remainder of this section, we explain each part of the model, by defining its dimensions, explaining their importance, and proposing characterization attributes (summarised in Table~\ref{tab:characterizationattributes}), in order to assess the fitness to each use perspective.
We define user as any of the stakeholders aiming to consume data from an open budget initiative.

{
\renewcommand{\arraystretch}{1.8}
\begin{table*}[t]
\caption{Model parts, dimensions and characterization attributes defined to characterize an Open Budget Initiative.}
\label{tab:characterizationattributes}
\begin{tabular}{llll}
\hline
\multicolumn{1}{c}{\textbf{Model Part}} & \multicolumn{1}{c}{\textbf{Dimension}}  & \multicolumn{1}{c}{\textbf{Characterization Attribute}}                                                                & \multicolumn{1}{c}{\textbf{Possible Values}}                                                      \\ \hline
\multirow{6}{*}{General}                & \multirow{2}{*}{Objective} & \ca{ca:objective}: Is the objective clearly stated?                                                         & Yes/No                                                                                            \\ \cline{3-4} 
                                        &                                         & \ca{ca:audience}: Is the intended audience defined?                                                         & Yes/No                                                                                            \\ \cline{2-4} 
                                        & \multirow{3}{*}{Content}     & \ca{ca:exclusive}: Are data exclusively on budget?                                                          & Yes/No                                                                                            \\ \cline{3-4} 
                                        &                                         & \ca{ca:source}: What is the source of data?                                                                 & Primary Source/Secondary Source                                                                        \\ \cline{3-4} 
                                        &                                         & \ca{ca:scope}: What is the scope covered by the strategy?                                                   & \begin{tabular}[c]{@{}l@{}}Country/Regional/Local\\ Transnational/Generic, and Legislative\end{tabular}             \\ \cline{2-4} 
                                        & Responsibility                & \ca{ca:responsible}: Who is responsible for the strategy?                                                   & Government/Society/Both                                                                           \\ \hline
\multirow{8}{*}{Publishing}             & \multirow{3}{*}{Data}                   & \ca{ca:measures}: What measures are available?                                                             & Budget/Spending/Revenues/Generic                                                                          \\ \cline{3-4} 
                                        &                                         & \ca{ca:dimensions}: What dimensions are available?                                                         & Time/Place/Payer/Payee/Category/Generic                                                                             \\ \cline{3-4} 
                                        &                                         & \ca{ca:grain}: What is the finest data granularity?                                                                   & Transaction/Aggregate/Generic                                                                              \\ \cline{2-4} 
                                        & Formats                                  & \ca{ca:formats}: Which formats are available?                                                                        & Five Stars of Open Data \\ \cline{2-4} 
                                        & Metadata                                & \ca{ca:metadata}: Are metadata  available?                                                                    & Yes/No                                                                                                   \\ \cline{2-4} 
                                        & Semantics                               & \ca{ca:semantics}: Is any ontology or vocabulary used?                                                      & Yes/No                                                                                       \\ \cline{2-4} 
                                        & Access                                 &  \ca{ca:access}: How are data made available?                                                              &  \begin{tabular}[c]{@{}l@{}}Catalogue/Raw Data/Querying\\ System/Stories/Infographics\end{tabular}                             \\ \cline{2-4} 
                                        & License                                 & \ca{ca:license}: Are data licensed?                                                                          & Yes/No                                                                                            \\ \hline
\multirow{2}{*}{Consumption}            & Usability                               & \ca{ca:tool}: What software tool is used?                                                                                                           & CKAN/OpenSpending/Other                                                                                                   \\ \cline{2-4} 
                                        & Feedback                                & \ca{ca:feedback}: Is it possible to give feedback over data?                                                & Comments/Data Request/Issue Reporting                                                                                           \\ \hline
\end{tabular}
\end{table*}
}

\subsection{General Aspects}
	\subsubsection{Objective}
Motivations to publish budget data, or generally open data, can be very diverse. 
In Section~\ref{sec:introduction}, we listed five common reasons for publishing budget data: transparency, participation, comparative analysis, efficiency and effectiveness assessment, and generating business value.
Defining the aimed audience is also important, since different user profiles require different approaches.
For example, in UP\ref{up:transp}, detailed data in machine readable formats is desirable, while for UP\ref{up:part}, human readable charts and tables are most suitable.
A SPARQL endpoint could better fit the needs of UP\ref{up:policy}.

		\begin{defn}The \textbf{Objective} dimension represents the motivations alleged for publishing budget data, including the definition of the intended audience.
		\end{defn}
		\noindent\textbf{Characterization attributes:}
We define as a characterization attribute: (i) whether an initiative states clearly its objective (CA\ref{ca:objective}), and (ii) whether the intended audience is explicitly defined (CA\ref{ca:audience}). 

	\subsubsection{Content}
	\label{dim:content}
Open budget initiatives are very heterogeneous regarding to the presented content. 
Data can refer to several administration levels (local, regional, national), and also to the different power instances (Executive, Legislative or Judiciary), according to the political system of each country.

		\begin{defn}The \textbf{Content} dimension has the objective of assessing the nature of the information contained in an open budget initiative.
		\end{defn}
		\noindent\textbf{Characterization attributes:}
The first important distinction we want to highlight is whether the initiative is exclusively for publishing budget data, or it contains other kinds of information (CA\ref{ca:exclusive}). 
Then, we also distinguish primary sources of data from applications working over data published by other initiatives, that is, secondary data (CA\ref{ca:source}). 
Finally, we assess the scope of the initiative~(CA\ref{ca:scope}), classifying it into local (1), regional (2), national (3) or transnational (4) range.
A special sign identifies initiatives focused only on the legislative power (L), considering that initiatives, normally exhibit general budget data.
We also consider that the scope can be generic (5), when the initiative allows publishers to display different datasets, referring to different scopes.

	\subsubsection{Responsibility}
Publishing budgetary data implies a great responsibility of people in charge of the initiative.
This kind of information is quite sensible, and mistakes can lead to severe consequences. 
Government, as supplier of primary data, may define specific sectors to be responsible for publishing budget data. 
In the US, responsibility is under the General Services Administration, while in UK, there is a Transparency and Open Data team under the Cabinet Office. 
In Brazil, administration is under the Ministry of Planning, Budget and Management.
Organization of civil society also play an important role by building applications over primary data, specially regarding UP\ref{up:part}.
In this case, responsibility lies in making the context clear, and simplifying as much as possible for data to be understood, but as little as possible to avoid misinterpretations.
		\begin{defn}The \textbf{Responsibility} dimension of an open budget initiative refers to the person(s) or organization(s) responsible for publishing the data, from operational tasks up to guaranteeing the authenticity of the provided information.
		\end{defn}
		\noindent\textbf{Characterization attributes:}
We define, as a characterization attribute, the distinction between data provided by governments and by society (CA\ref{ca:responsible}).
We also consider the possibility of a joint government/society partnership.

\subsection{Publishing}
Actions to be taken referring to these dimensions are expected from data publishers, supposedly influenced by data consumers.

	\subsubsection{Data}
While the \emph{Content} dimension (\ref{dim:content}) aimed to deal with general aspects related to the content of an open budget initiative, the \emph{Data} dimension focuses on specific aspects.

		\begin{defn}The \textbf{Data} dimension represents specific aspects of the data content and determines what kind of information is possible to be extracted from an open budget initiative.
		\end{defn}
		\noindent\textbf{Characterization attributes:}
In order to characterize the data content, we define three characterization attributes: 
(i) \emph{Measures}, i.e., the types of represented quantities, which can be budget, spendings and/or revenues (CA\ref{ca:measures}); 
(ii) \emph{Dimensions}, i.e., how the measures are qualified, which can be time, space and/or other categories (CA\ref{ca:dimensions}); and 
(iii) \emph{Granularity}, i.e., the finest level of detail available: transaction or aggregate (CA\ref{ca:grain}). 
For all CAs in this dimension, we also accept the generic value, when the options are not predefined and several datasets in the same initiative present different settings.

	\subsubsection{Formats}
When data are offered for download, the format in which they are encoded plays a very important role. For UP\ref{up:transp}, data in machine readable formats are crucial. 
For UP\ref{up:policy}, unique identification of entities and relations is also very important. 
The semantic resources generated by open budget initiatives can be instantly ready for reuse, when resources follow Linked Open Data (LOD) principles and guidelines~\cite{_linked_2011}. 
In this case, all URIs must be resolvable and dereferenceable. These resources shall be accessible via a SPARQL\footnote{SPARQL is a set of specifications to query and manipulate RDF graph content on the Web. More on \url{http://www.w3.org/sparql/}.} endpoint, or by directly resolving resource URIs.
The latter must return either an RDF representation of the resource, or a more eye-friendly HTML visualisation, according to the negotiated content-type. 
It is of utmost importance that the resources are available on a stable server. 
This is also important as these resources could be linked to others in the LOD cloud. 
The resulting data, which will be in a standard interoperable format (RDF), will be fully compliant with the statement for best practices given by the G8 Science Ministers~\cite{RoyalSociety2013}: "Data should be easily discoverable, accessible, assessable, intelligible, useable, and wherever possible interoperable to specific quality standards". 
Due to LOD being a widespread initiative, existing tools can be exploited and used in order to reuse datasets.

		\begin{defn}The \textbf{Formats} dimension represents the type of formats in which downloadable data are offered by an open budget initiative.
		\end{defn}
		\noindent\textbf{Characterization attributes:}
Here, we adopt the well established open data five stars model\footnote{http://5stardata.info/} as characterization attribute (CA\ref{ca:formats}).

	\subsubsection{Metadata}
Adequate metadata are fundamental for providing complementary information about the context in which data are immersed. 
Information such as dataset author, published date and last update, formats and license are usually the basic metadata. 
Another useful class of metadata is provenance. 
Provenance metadata describe the transformations applied to the dataset, and can also explain the process through which each data item was generated.

		\begin{defn}The \textbf{Metadata} dimension refers to the availability of descriptors associated to the provided datasets.
		\end{defn}
		\noindent\textbf{Characterization attributes:}
As a characterization attribute, we check for the existence of metadata in an open budget initiative (CA\ref{ca:metadata}).

	\subsubsection{Semantics}
In order to be correctly interpreted, data must be contextualized to avoid problems that emerge from terminology ambiguity or lack of agreement. 
Without post-hoc unification the data may be difficult to understand, as their users may need to familiarize themselves with different terminologies for each dataset.
Having a single data format may solve structural heterogeneity, at the expenditure of the cost of introducing yet another format bridging the others. 
A more complex issue refers to semantic heterogeneity, which may be addressed by simpler solutions based on vocabularies to more comprehensive approaches based on ontologies. 
It is therefore fundamental not to multiply the competing approaches for modelling public budgets and spending data, but rather build on previous work, such as~\cite{OpenSpending}, and align divergent approaches using links and semantic relations.

The current repositories of public finance data, such as OpenSpending.org, serve well as data catalogues, in which each dataset exists more or less in isolation as a separate black box. 
The absence of links and explicit semantics forms a barrier to automated processing, combining, and joining datasets of distinct origin. 
In the context of such tasks, applying linked data and semantic web technologies offers greater data interoperability. 

Nevertheless, perhaps the most important are the benefits of linked data for improving data interpretation. 
The key to such improvement comes from the recognition that measures in public budget and spending data are relative.
If there is no way to compare them and put them into context, it is difficult to make sense of the data.
Putting money into a wider context, on which it was spent, helps to perform meaningful analyses and find comprehensible "stories" in data. 
The context may be provided by linked datasets, such as population statistics. 
Added links to external data can link public finance with the LOD cloud, offering many ways to view data given different contextualising information, such as economic indicators or demographic statistics. 
Ultimately, a key goal of the proposed data model is to enable better comprehension of public finance data. 

For UP\ref{up:policy}, following semantic standards is mandatory.
Even though budget data tends to be very heterogeneous, especially between different countries, some common points can be found, for example spending categories (Health, Education, Debt Services) or international companies. 
Budget ontologies regarding specific countries have been developed~\cite{Santos2012, Reynolds2010}, and even an international effort is in course~\cite{OpenSpending}. 
Although not providing immediate linking possibilities, following standards as the Special Data Dissemination Standard~\cite{InternationalMonetaryFund2013} helps to make data comparable.

		\begin{defn}The \textbf{Semantics} dimension refers to the support of any terminological complementary resource that allows a better understanding of the data domain concepts.
		\end{defn}

		\noindent\textbf{Characterization attributes:}
We define the \emph{Semantics} characterization attribute as a boolean value, that indicates the presence of standardized vocabularies or ontologies (CA\ref{ca:semantics}) in the open budget initiative.

	\subsubsection{Access}
The simplest way of publishing budget information is by offering data for download, which can be done in several formats. 
However, in UP\ref{up:part}, interactive charts, maps or infographics are more useful than downloadable datasets, even if this might not be considered open data in the strict sense.
Thus, the \emph{Access} dimension aims to check the adequacy between the desired audience and the way data are offered.

		\begin{defn}The \textbf{Access} dimension refers to how the initiative presents budget data to its audience.
		\end{defn}
		\noindent\textbf{Characterization attributes:}
\emph{Data Access} is a characterization attribute (CA\ref{ca:access}) which can be assigned as:
\begin{itemize}
	\item Downloadable data, Linked Data/SPARQL endpoint;
	\item Data and metadata catalogue;
	\item Exploration by Tables;
	\item Visualization by Charts, Maps, Comparison; and/or
	\item Stories
\end{itemize}

	\subsubsection{Licensing}
Licensing is a fundamental issue for data reuse. 
In UP\ref{up:transp}, some kinds of use can be hindered by the absence of adequate licensing, for example, the development of derived applications. 
Currently, 3 types of general licenses for open data are available\footnote{See more at: \url{http://opendatacommons.org/licenses/}.}: Public Domain Dedication and License (PDDL), Attribution License (ODC-By), and Open Database License (ODC-ODbL). Some governments developed their own open data licenses, for example, Germany\footnote{\url{https://www.govdata.de/dl-de/by-1-0}} and UK\footnote{\url{http://www.nationalarchives.gov.uk/doc/open-government-licence/version/3/}}

		\begin{defn}The \textbf{Licensing} dimension assesses the legal status of data available in an open budget initiative.
		\end{defn}
		\noindent\textbf{Characterization attributes:}
We define a boolean characterization attribute to describe the existence of a license (CA\ref{ca:license}) on data published by an open budget initiative.

\subsection{Consumption}
The justification and characterization attributes identified in this paper aimed at the success of the use perspectives.
In this part, we detail specific issues related to actions to be taken by the users, when interfacing with budget data. 

	\subsubsection{Usability/Design}

A good set of visualisations, which are self explanatory and easy to understand, certainly can improve usage of an open budget initiative.
Interactive visualisations and infographics can also enable a stakeholder to focus on a particular aspect of the data.
In \cite{Walker2010}, impacts of usability and design issues are discussed. The experiments showed how improvements on design led to better results with users.

Several aspects of this dimension overlap with dimensions of the \emph{Publishing} part.
Particularly, different ways of accessing data (\emph{Access} dimension) heavily impact usability, and exporting data in different formats (\emph{Formats} dimension), such as CSV, XML or RDB is also important to encourage the reuse of data.
Thus, the way data are published can enable stakeholders to get the most out of the open data. 
		\begin{defn}The \textbf{Usability/Design} dimension verifies if the initiative interface is suitable to the requirements of the use perspective.
		\end{defn}
		\noindent\textbf{Characterization attributes:}
The complexity of analysing user interfaces surpasses the scope of this paper. 
Nevertheless, we define a characterization attribute related to the software tool used by the initiative (CA\ref{ca:tool}), understanding that the tool behind the initiative plays an important role on the usability. 
Possible values are the two major open source software tools available for publishing open data: \href{http://github.com/openspending/}{OpenSpending} and \href{http://ckan.org}{CKAN}.

	\subsubsection{Feedback}
In order to enable the collaboration between the public sector administration and the other stakeholders, open budget initiatives have to provide means to discuss and give feedback on the provided data. 
This feedback might be provided to the public administrators either as comments or as a set of recommendations.
For example, NGOs could give feedback on what should the budget focus on their practice area. 
Ideally, this communication process should be transparent, that is, feedback and recommendations given to public administrators should be publicly available and any changes, resulting from the feedback, should be recorded.

The importance of stimulating user engagement on open data initiatives through feedback and collaboration has been stressed by the Five Stars of Open Data Engagement model~\cite{Davies2012}.
This model justifies the necessity of data being demand driven, contextualized, and collaborative. 
The conversation around data is also pointed as a strategy to engage users.
According to this model, data should be regarded as a common resource, what enforces the necessity of collaboration. 
The lack of collaboration has been listed by~\cite{Zuiderwijk2014} as one of the main factors hindering the development of open data policies.

To enable collaboration, tools to allow feedback on budget allocations and specific expenditure transactions should be provided to stakeholders. 
Public administrations must have the instruments to receive and effectively manage this feedback, enabling greater degrees of active citizen involvement and participation.

		\begin{defn}The \textbf{Feedback} dimension represents the user's capacity to collaborate in data publishing and express her/his opinion.
		\end{defn}
		\noindent\textbf{Characterization attributes:}
Although this point requires a deeper analysis, we noticed that many open budget initiatives do not present any feedback support. 
Thus, we define one basic binary characterization attribute which is the existence of feedback mechanism~(CA\ref{ca:feedback}). 
We check if it is possible to: 
(i) comment on data;
(ii) submit a new data request; and 
(iii) report issues noticed in data analysis.

\section{Analysis of Open Budget Data initiatives}
\label{sec:analysis} 
In this section, we describe the application of the model to a number of open budget initiatives. 
The goal of this evaluation is not to be extensive or to achieve statistical significance, but rather to test the model, to discover its potentials and limitations, and to gain some intuition on the domain.
Results are shown in Table~\ref{tab:results}, and data can be accessed at \url{http://bit.ly/1FNThhH}.

The 23 initiatives were chosen considering a balance between primary (11) and secondary (12) sources (CA\ref{ca:source}). 
The sample also contains at least five initiatives strongly related to each use perspective, and considers initiatives from 6 countries plus the European Union, presented in five different idioms. Some of the analysed initiatives are listed on the \emph{Map of Spending Projects}\footnote{Available at \url{http://community.openspending.org/map-of-spending-projects/}.}.

All primary sources are maintained by the government, and most of the secondary ones are society driven. 
Among them, two initiatives were identified as maintained in partnership between government and society organizations (CA\ref{ca:responsible}). 
Initiatives generally display their objectives (22 - CA\ref{ca:objective}), but only 11 explicitly mention their intended audience (CA\ref{ca:audience}). 
Also, almost all initiatives offer data for download (18), which favours UP\ref{up:transp}, and more than half of them (13) make visualization available, favouring UP\ref{up:part}.

Even considering the low number of initiatives evaluated, two outcomes drew the attention, regarding feedback and semantics. 
Commenting on data is allowed only in three initiatives, and the same number (but not the same ones) offers a data request form.
No reporting issues mechanisms were found, revealing a strong absence of feedback possibilities (CA\ref{ca:feedback}).

The lack of semantics support (only three offered it - CA\ref{ca:semantics}), or linkable data (again, only three had it - CA\ref{ca:formats}) also may point that UP\ref{up:policy} is still far from reality. 
Ten initiatives use categories for the datasets, which at least facilitate some form of comparisons.

Regarding the use perspectives, we can state:

\hspace{.8cm}

\noindent\textbf{UP\ref{up:transp} - Transparency:}
The main requirements for this use perspective -- data on transaction level, machine readable formats, aggregation levels -- were accomplished by most of the open budget initiatives. 
However, much work is still to be done concerning the feedback handling.
We can say that, for most of the analysed cases, stakeholders interested in auditing government and in translating data into more accessible formats are partially satisfied.

\hspace{.8cm}

\noindent \textbf{UP\ref{up:part} - Participation:}
The requirements set for this use perspective enforced human readable formats, that allows citizens without deep budget knowledge to understand data and to participate in discussions.
Slightly more than half of the initiatives present graphics, which can help quick insights over data.
Only three initiatives offer maps to visualize budget data, what is coherent to the low number of initiatives that include the location dimension (eight).
Another aspect emphasized in this use perspective was the usability and design. 
Considering the already mentioned limitations on assessing this issue, we noticed that ten initiatives use standard open source software tools. 
Although this is not the most relevant factor regarding usability, the use of standard tools favours users dealing with several open budget initiatives.
Moreover, as open source tools, the more initiatives using these tools, the better they can be developed.

\hspace{.8cm}

\noindent \textbf{UP\ref{up:policy} - Policy Making:}
The main requirements in this perspective were the use of common classifications, vocabularies and ontologies, and the possibility of linking data with other databases.
As already mentioned, semantics support was mostly absent. Comparison tools, also important in this case, were found only in three of the initiatives. 
Thus, this use perspective is still far from being realised in most of the analysed initiatives. 
All these indicate that working on standard terminologies and common conceptualizations as suggested by OpenSpending~\cite{OpenSpending} is highly desirable.

\section{Conclusions}
\label{sec:conclusions} 
In this paper, we presented a model to analyse open budget initiatives, including dimensions and assessable characterization attributes. 
The model covers, at this stage, General Aspects, Publishing and Consumption dimensions of these initiatives. 
Initial testing of the model and analysis of 23 open budget initiatives revealed that attention has to be given on feedback, semantics support and linking possibilities.

Future research includes adding the Context dimensions in the model, and developing characterization attributes for it. 
We intend also to extend the assessment to other initiatives, as well as to conduct a more detailed analysis of some of their characteristics.
As the results pointed a very weak performance on the Feedback dimension, we aim to further explore the Consumption part, in order to propose solutions that can contribute on this issue. 
The Usability dimension also needs more consistent characteristic attributes.
The use of standard vocabularies or ontologies and linkable data formats also deserves attention.

Regarding the use perspectives, we conclude that transparency requisites were mostly accomplished by the analysed initiatives.
Participation, in turn, is still not heavily supported, while tools for comparing budget data by policy makers are still far from reality.

It has to be noticed that materializing transparency is far more complex than just publishing budget data through software tools.
Several political issues are related to data publishing, as well as deep data literacy questions are involved in the usage of open budget initiatives.
Gurstein~\cite{Gurstein2011} alerts about the emergence of a "data divide", a parallel concept to the digital divide, distinguishing people "who have access to data which could have significance in their daily lives and those who don't".
Thus, transparency policies can not be implemented without actions to foster digital inclusion, and why not, "data inclusion".

Software tools are, although indispensable, just part of the process. 
With this research effort, we aim to enrich the existing knowledge on open budget, and help to set the basis for developing tools and procedures that, together with other actions, may result in real benefits of fiscal transparency to the society. 

\newcommand{\rot}[1]{\begin{sideways}#1\end{sideways}}

\section{Acknowledgments}
A. Tygel is supported by CAPES/PDSE grant 99999.008268/2014-02. M.L.M.Campos is partially supported by CNPq--Brazil.
Also, this work is supported in part by the European Commission under the Seventh Framework Program FP7/2007-2013 (\emph{LinDA} -- GA ICT-610565).

{
\renewcommand{\arraystretch}{1.8}
\begin{sidewaystable*}[h]
\caption{Results of the application of the model on some open budget initiatives. 
In CA\ref{ca:scope}: (1) local; (2) regional; (3) national; (4) transnational; (5) generic; and (L) legislative budget. 
In CA\ref{ca:responsible}: (Gov) Government; and (Soc) Society.
In CA\ref{ca:grain}: (Tr) transaction; (Ag) aggregate; and (Ge)generic.
In CA\ref{ca:formats}, N/A means not applicable, when there is no data for download.
In CA\ref{ca:tool}: (OS) \href{http://github.com/openspending/}{OpenSpending}; (CK) \href{http://ckan.org}{CKAN}; and other used specific non open source software.
This table can be accessed online at \url{http://bit.ly/1FNThhH}.
}
\label{tab:results}
\scriptsize
\begin{tabular}{|l|l|l|l|l|l|l|l|l|l|l|l|l|l|l|l|l|l|l|l|l|l|l|l|l|l|l|l|l|l|l|l|l|l|l|}
\hline
\multicolumn{1}{|c|}{\multirow{4}{*}{ID}} & \multicolumn{6}{c|}{General Aspects}                                                                                                                                                                                                                                                                                                                                             & \multicolumn{22}{c|}{Publishing}                                                                                                                                                                                                                                                                                                                                                                                                                                                                                                                                                                                                                                                                                                                                                                                                                                                                                                                                                                          & \multicolumn{4}{c|}{Consumption}                                                                                                                                                                    & \multicolumn{1}{c|}{\multirow{4}{*}{\rot{Use Perspective}}} & \multicolumn{1}{c|}{\multirow{4}{*}{\rot{Language}}} \\ \cline{2-33}
\multicolumn{1}{|c|}{}                    & \multicolumn{3}{c|}{\rot{Objective}}                                                                                                                                                    & \multicolumn{2}{c|}{\rot{Content}}                                                                                 & \multicolumn{1}{c|}{\rot{Responsibility}}                      & \multicolumn{11}{c|}{Data}                                                                                                                                                                                                                                                                                                                                                                                                                             & \multicolumn{1}{c|}{\rot{Formats}}                        & \multicolumn{1}{c|}{\rot{Metadata}}                       & \multicolumn{1}{c|}{\rot{Semantics}}                       & \multicolumn{7}{c|}{Acces}                                                                                                                                                                                                                                                               & \multicolumn{1}{c|}{\rot{License}}                       & \multicolumn{1}{c|}{\rot{Usability}}                         & \multicolumn{3}{c|}{\rot{Feedback}}                                                                                            & \multicolumn{1}{c|}{}                    & \multicolumn{1}{c|}{}                       \\ \cline{2-33}
\multicolumn{1}{|c|}{}                    & \multicolumn{1}{c|}{\multirow{2}{*}{CA\ref{ca:objective}}} & \multicolumn{1}{c|}{\multirow{2}{*}{CA\ref{ca:audience}}} & \multicolumn{1}{c|}{\multirow{2}{*}{CA\ref{ca:exclusive}}} & \multicolumn{1}{c|}{\multirow{2}{*}{CA\ref{ca:source}}} & \multicolumn{1}{c|}{\multirow{2}{*}{CA\ref{ca:scope}}} & \multicolumn{1}{c|}{\multirow{2}{*}{CA\ref{ca:responsible}}} & \multicolumn{4}{c|}{CA\ref{ca:measures}}                                                                                                                 & \multicolumn{6}{c|}{CA\ref{ca:dimensions}}                                                                                                                                                                                    & \multicolumn{1}{c|}{\multirow{2}{*}{CA\ref{ca:grain}}} & \multicolumn{1}{c|}{\multirow{2}{*}{CA\ref{ca:formats}}} & \multicolumn{1}{c|}{\multirow{2}{*}{CA\ref{ca:metadata}}} & \multicolumn{1}{c|}{\multirow{2}{*}{CA\ref{ca:semantics}}} & \multicolumn{7}{c|}{CA\ref{ca:access}}                                                                                                                                                                                                                                                 & \multicolumn{1}{c|}{\multirow{2}{*}{CA\ref{ca:license}}} & \multicolumn{1}{c|}{\multirow{2}{*}{CA\ref{ca:tool}}} & \multicolumn{3}{c|}{CA\ref{ca:feedback}}                                                                                       & \multicolumn{1}{c|}{}                    & \multicolumn{1}{c|}{}                       \\ \cline{8-17} \cline{22-28} \cline{31-33}
\multicolumn{1}{|c|}{}                    & \multicolumn{1}{c|}{}                                        & \multicolumn{1}{c|}{}                                       & \multicolumn{1}{c|}{}                                        & \multicolumn{1}{c|}{}                                     & \multicolumn{1}{c|}{}                                    & \multicolumn{1}{c|}{}                                         & \multicolumn{1}{c|}{\rot{Generic}} & \multicolumn{1}{c|}{\rot{Budget}} & \multicolumn{1}{c|}{\rot{Spending}} & \multicolumn{1}{c|}{\rot{Revenue}} & \multicolumn{1}{c|}{\rot{Generic}} & \multicolumn{1}{c|}{\rot{Time}} & \multicolumn{1}{c|}{\rot{Place}} & \multicolumn{1}{c|}{\rot{Payer}} & \multicolumn{1}{c|}{\rot{Payee}} & \multicolumn{1}{c|}{\rot{Category}} & \multicolumn{1}{c|}{}                                    & \multicolumn{1}{c|}{}                                       & \multicolumn{1}{c|}{}                                       & \multicolumn{1}{c|}{}                                        & \multicolumn{1}{c|}{\rot{Downloadable Data~~~}} & \multicolumn{1}{c|}{\rot{Catalogue}} & \multicolumn{1}{c|}{\rot{Table}} & \multicolumn{1}{c|}{\rot{Graphics}} & \multicolumn{1}{c|}{\rot{Map}} & \multicolumn{1}{c|}{\rot{Comparison}} & \multicolumn{1}{c|}{\rot{Stories}} & \multicolumn{1}{c|}{}                                      & \multicolumn{1}{c|}{}                                          & \multicolumn{1}{c|}{\rot{Comments}} & \multicolumn{1}{c|}{\rot{Data Request}} & \multicolumn{1}{c|}{\rot{Issue Reporting}} & \multicolumn{1}{c|}{}                    & \multicolumn{1}{c|}{}                       \\ \hline
1                                         & Yes                                                          & Yes                                                         & Yes                                                          & Sec                                                       & 1                                                        & Both                                                          &                                      & x                                   &                                       & x                                    &                                      & x                                 & x                                  & x                                  &                                    & x                                     & Ag                                                       & 3                                                           & No                                                          & No                                                           & x                                              &                                        & x                                  & x                                     & x                                &                                         &                                      & No                                                         & OS                                                             &                                       &                                           &                                              & UP1, UP2                                 & DE                                          \\ \hline
2                                         & Yes                                                          & Yes                                                         & No                                                           & Sec                                                       & 5                                                        & Soc                                                           &                                      &                                     & x                                     &                                      & x                                    & x                                  & x                                   & x                                   &  x                                  &                                       & Ag                                                       & 1                                                           & No                                                          & No                                                           &                                                &                                        &                                    &                                       &                                  & x                                       & x                                    & No                                                         &                                                                &                                       &                                           &                                              & UP1, UP2                                 & EN                                          \\ \hline
3                                         & Yes                                                          & No                                                          & Yes                                                          & Sec                                                       & 5                                                        & Soc                                                           & x                                    & x                                    &  x                                     &                                      & x                                    & x                                  & x                                   & x                                   &   x                                 &  x                                     & Tr                                                       & 3                                                           & Yes                                                         & No                                                           & x                                              & x                                      & x                                  & x                                     &                                  &                                         &                                      & Yes                                                        & OS                                                             &                                       &                                           &                                              & UP3                                      & EN                                          \\ \hline
4                                         & Yes                                                          & Yes                                                         & No                                                           & Sec                                                       & 1,2,3                                                    & Soc                                                           & x                                    & x                                    & x                                      &  x                                    &  x                                    &   x                                & x                                   &  x                                  &    x                                &  x                                     & Ge                                                       & 1--3                                                        & Yes                                                         & No                                                           & x                                              & x                                      &                                    &                                       &                                  &                                         &                                      & Yes                                                        & CK                                                             & x                                     &                                           &                                              & UP1, UP2                                 & DE                                          \\ \hline
5                                         & Yes                                                          & No                                                          & No                                                           & Prim                                                      & 1                                                        & Gov                                                           & x                                    &                                     &                                       &                                      & x                                    &                                   &                                    &                                    &                                    &                                       & Ge                                                       & 1--3                                                        & Yes                                                         & No                                                           & x                                              & x                                      &                                    &                                       &                                  &                                         &                                      & Yes                                                        & CK                                                             & x                                     &                                           &                                              & UP1,UP2                                  & DE                                          \\ \hline
6                                         & Yes                                                          & No                                                          & No                                                           & Prim                                                      & 1                                                        & Gov                                                           &   x                                   &                                     & x                                     &                                      &                                      & x                                 & x                                  & x                                  & x                                  & x                                     & Ge                                                       & 3                                                           & Yes                                                         & No                                                           & x                                              & x                                      & x                                  &                                       & x                                &                                         &                                      & Yes                                                        & CK                                                             & x                                     &                                           &                                              & UP1,UP2                                  & EN                                          \\ \hline
7                                         & Yes                                                          & No                                                          & No                                                           & Prim                                                      & 1                                                        & Gov                                                           & x                                    &                                     &                                       &                                      & x                                    &                                   &                                    &                                    &                                    &                                       & Ge                                                       & 3                                                           & Yes                                                         & No                                                           & x                                              & x                                      &                                    &                                       &                                  &                                         &                                      & No                                                         & CK                                                             &                                       & x                                         &                                              & UP1                                      & EN                                          \\ \hline
8                                         & Yes                                                          & Yes                                                         & No                                                           & Prim                                                      & 1                                                        & Gov                                                           & x                                    & x                                   &                                       &                                      & x                                    &                                   &                                    &                                    &                                    &                                       & Ge                                                       & 1--5                                                        & Yes                                                         & Yes                                                          & x                                              & x                                      & x                                  & x                                     &                                  &                                         &                                      & Yes                                                        & CK                                                             &                                       &                                           &                                              & UP1, UP2                                 & PT                                          \\ \hline
9                                         & Yes                                                          & No                                                          & Yes                                                          & Sec                                                       & 1,2,3 L                                                  & Soc                                                           &                                      &                                     & x                                     &                                      &                                      & x                                 &                                    & x                                  & x                                  & x                                     & Tr                                                       & N/A                                                         & No                                                          & No                                                           &                                                &                                        & x                                  & x                                     &                                  &                                         &                                      & Yes                                                        &                                                                &                                       &                                           &                                              & UP2                                      & PT                                          \\ \hline
10                                        & Yes                                                          & Yes                                                         & Yes                                                          & Sec                                                       & 5                                                        & Soc                                                           &                                      &                                     & x                                     &                                      &                                      &                                   &                                    &                                    &                                    &                                       & Tr                                                       & 5                                                           & Yes                                                         & Yes                                                          & x                                              &                                        & x                                  & x                                     &                                  & x                                       &                                      & No                                                         &                                                                &                                       &                                           &                                              & UP3                                      & EN                                          \\ \hline
11                                        & Yes                                                          & No                                                          & Yes                                                          & Prim                                                      & 3                                                        & Gov                                                           &                                      & x                                   & x                                     & x                                    &                                      & x                                 &                                    & x                                  & x                                  & x                                     & Tr                                                       & 3                                                           & No                                                          & No                                                           & x                                              &                                        & x                                  & x                                     &                                  &                                         &                                      & No                                                         &                                                                &                                       & x                                         &                                              & UP1, UP2                                 & PT                                          \\ \hline
12                                        & Yes                                                          & No                                                          & Yes                                                          & Sec                                                       & 1,2,3                                                    & Gov                                                           &                                      &                                     & x                                     &                                      &                                      & x                                 & x                                  & x                                  & x                                  & x                                     & Tr                                                       & 3                                                           & Yes                                                         & No                                                           & x                                              &                                        & x                                  & x                                     & x                                &                                         &                                      & No                                                         &                                                                &                                       &                                           &                                              & UP1,UP2                                  & EN                                          \\ \hline
13                                        & Yes                                                          & Yes                                                         & Yes                                                          & Prim                                                      & 1                                                        & Gov                                                           &                                      &                                     & x                                     & x                                    &                                      & x                                 &                                    & x                                  & x                                  & x                                     & Tr                                                       & 3                                                           & No                                                          & No                                                           & x                                              &                                        & x                                  &                                       &                                  &                                         &                                      & No                                                         &                                                                &                                       &                                           &                                              & UP1,UP2                                  & PT                                          \\ \hline
14                                        & Yes                                                          & No                                                          & Yes                                                          & Prim                                                      & 2                                                        & Gov                                                           &                                      & x                                   & x                                     & x                                    &                                      & x                                 &                                    & x                                  & x                                  & x                                     & Tr                                                       & 3                                                           & No                                                          & No                                                           & x                                              &                                        & x                                  & x                                     &                                  &                                         &                                      & No                                                         &                                                                &                                       &                                           &                                              & UP1, UP2                                 & PT                                          \\ \hline
15                                        & Yes                                                          & Yes                                                         & Yes                                                          & Sec                                                       & 1                                                        & Soc                                                           &                                      & x                                   &                                       &                                      &                                      & x                                 &                                    & x                                  &                                    & x                                     & Ag                                                       & N/A                                                         & No                                                          & No                                                           &                                                &                                        & x                                  & x                                     &                                  &                                         &                                      & No                                                         & OS                                                             &                                       &                                           &                                              & UP2                                      & PT                                          \\ \hline
16                                        & Yes                                                          & Yes                                                         & Yes                                                          & Prim                                                      & 1,2,3                                                    & Gov                                                           &                                      & x                                   & x                                     & x                                    &                                      & x                                 & x                                  & x                                  & x                                  & x                                     & Tr                                                       & 3                                                           & Yes                                                         & No                                                           & x                                              &                                        & x                                  &                                       &                                  &                                         &                                      & No                                                         &                                                                &                                       &                                           &                                              & UP1                                      & PT                                          \\ \hline
17                                        & Yes                                                          & Yes                                                         & Yes                                                          & Sec                                                       & 3                                                        & Soc                                                           &                                      & x                                   & x                                     & x                                    &                                      & x                                  & x                                   & x                                   &                                    &                                      x & Ag                                                       & N/A                                                           & Yes                                                         & No                                                            &                                                &                                        & x                                  & x                                     &                                  & x                                       & x                                    & Yes                                                        &                                                                &                                       &                                           &                                              & UP3                                      & IT                                          \\ \hline
18                                        & Yes                                                          & Yes                                                         & Yes                                                          & Sec                                                       & 4                                                        & Both                                                          &                                      &                                     & x                                     &                                      &                                      & x                                 & x                                  & x                                  & x                                  & x                                     & Tr                                                       & 3                                                           & Yes                                                         & No                                                           & x                                              &                                        &                                    &                                       &                                  &                                         &                                      & Yes                                                        &                                                                &                                       &                                           &                                              & UP1, UP3                                 & EN                                          \\ \hline
19                                        & Yes                                                          & Yes                                                         & Yes                                                          & Sec                                                       & 4                                                        & Soc                                                           &                                      &                                     & x                                     &                                      &                                      & x                                 & x                                  & x                                  & x                                  & x                                     & Tr                                                       & N/A                                                         & Yes                                                         & No                                                           & x                                              &                                        & x                                  &                                       &                                  & x                                       &                                      & No                                                         & OS                                                             &                                       &                                           &                                              & UP3                                      & EN                                          \\ \hline
20                                        & No                                                           & No                                                          & Yes                                                          & Prim                                                      & 1                                                        & Gov                                                           &                                      & x                                   &  x                                     &                                      &                                      & x                                 & x                                  & x                                  & x                                  & x                                     & Tr                                                       & 5                                                           & Yes                                                         & Yes                                                          & x                                               &                                        &                                    &                                       &                                  &                                         &                                      & No                                                         &                                                                &                                       &                                           &                                              & UP1                                      & PT                                          \\ \hline
21                                        & Yes                                                          & No                                                          & No                                                           & Prim                                                      & 3                                                        & Gov                                                           & x                                    & x                                   & x                                     & x                                    & x                                    & x                                 &                                    & x                                  & x                                  & x                                     & Tr                                                       & 3                                                           & Yes                                                         & No                                                           & x                                              &                                        & x                                  & x                                     &                                  &                                         &                                      & Yes                                                        &                                                                &                                       &                                           &                                              & UP1,UP2                                  & RU                                          \\ \hline
22                                        & Yes                                                          & Yes                                                         & No                                                           & Prim                                                      & 3                                                        & Gov                                                           & x                                    & x                                   & x                                     & x                                    & x                                    & x                                 &                                    & x                                  & x                                  & x                                     & Tr                                                       & 3                                                           & Yes                                                         & No                                                           & x                                              &                                        & x                                  & x                                     &                                  &                                         &                                      & Yes                                                        &                                                                &                                       & x                                         &                                              & UP1,UP2                                  & RU                                          \\ \hline
23                                        & Yes                                                          & No                                                          & Yes                                                          & Sec                                                       & 2                                                        & Soc                                                           & x                                    & x                                   & x                                     &                                      & x                                    & x                                 &                                    & x                                  & x                                  & x                                     & Tr                                                       & 3                                                           & Yes                                                         & No                                                           & x                                              &                                        & x                                  & x                                     &                                  &                                         &                                      & Yes                                                        & OS                                                             &                                       &                                           &                                              & UP1,UP2                                  & RU                                          \\ \hline
\end{tabular}
\end{sidewaystable*}
}
%
\bibliographystyle{abbrv}
\bibliography{bibliography,../../bib/aksw}  
%
%
\end{document}